\documentstyle[aps]{revtex}

\newcommand{\be}{\begin{equation}}
\newcommand{\ee}{\end{equation}}
\newcommand{\bea}{\begin{eqnarray}}
\newcommand{\eea}{\end{eqnarray}}
\newcommand{\nl}{\\ \nonumber}

\newcommand{\order}{{\cal O}}

\newcommand{\sig}{\sigma_e\cdot\sigma_\mu}

\begin{document}

\title{New Value of $m_\mu/m_e$ from Muonium Hyperfine Splitting}

\date{\today}
\author{Richard J.\ Hill \cite{email-rjh} }

\address{Newman Laboratory of Nuclear Studies, Cornell University\\
Ithaca, New York 14853}
\maketitle

\begin{abstract}
The complete contribution to the muonium hyperfine splitting
of relative order 
$\alpha^3(m_e/m_\mu)\ln\alpha$ is calculated. The result amounts
to $0.013\,{\rm kHZ}$, much smaller than suggested by a previous 
estimate, and leads to a $2\sigma$ upward shift 
of the most precise value for the muon-electron mass ratio,
with the error reduced by 
approximately $30 \%$.
Analogous contributions are calculated for the positronium 
hyperfine splitting:
$(217/90-17\ln{2}/3)m_e(\alpha^7/\pi)\ln\alpha^{-1} \approx -0.32\,{\rm MHz}$;
the remaining theoretical uncertainty is well below experimental error,
leaving discrepancies of $2.5\sigma$ and $3.5\sigma$ with the two most precise 
measurements. 
\end{abstract}

\vspace{2mm}

Precise measurement of the ground-state muonium ($\mu^+ e^-$)
hyperfine-splitting (HFS), together with the corresponding
theoretical analysis, provides a stringent test of bound 
state theory in Quantum Electrodynamics (QED), and
allows a precise determination
of the fundamental physical constants $m_\mu/m_e$ and $\alpha$.
The most precise measurement gives~\cite{Liu}:
\be
\label{eq:expt}
\Delta\nu({\rm Mu})_{\rm expt.} = 4 \,463\, 302.765(53)\, {\rm kHz}
\,\,[1.2\times 10^{-8}].
\ee
The corresponding theoretical prediction can be expressed as a series expansion
in small parameters $\alpha\approx 1/137$ and $m_e/m_\mu\approx 1/207$; 
terms involving logarithms, $\ln\alpha^{-1}\approx\ln(m_\mu/m_e)\approx 5$, 
also appear.  At leading order in $\alpha$, 
the splitting is given by the Fermi energy~\cite{Z}: 
\be
E_F = h\Delta\nu_F =\frac{16}{3}(hc)R_\infty Z^4\alpha^2\frac{m_e}{m_\mu}\left[1+\frac{m_e}{m_\mu}\right]^{-3}.
\ee
The complete splitting 
can be broken into the sum of terms~\cite{Mohr}\cite{reviews}:
\be
\Delta\nu({\rm Mu})_{\rm theory} = \Delta\nu_D + \Delta\nu_{\rm rad}
	+\Delta\nu_{\rm rec} + \Delta\nu_{\rm r-r} + \Delta\nu_{\rm weak}
	+\Delta\nu_{\rm had}.
\ee
Here D stands for Dirac, or relativistic corrections, while   the other
terms are from radiative, recoil, radiative-recoil, weak and 
hadronic contributions.  

Currently, theory is limited by uncalculated or imprecisely-known 
terms in $\Delta\nu_{\rm rec}$ and $\Delta\nu_{\rm r-r}$
of order $E_F\alpha^3(m_e/m_\mu)$, some of which are 
enhanced by logarithmic factors; see Table~I.  
This paper presents a calculation of terms of order
$E_F\alpha^3(m_e/m_\mu)\ln\alpha$, with results:  
\bea
\label{eq:rec}
\delta(\Delta\nu_{\rm rec}) &=&
E_F\frac{(Z\alpha)^3}{\pi}\frac{m_e}{m_\mu}\ln(Z\alpha)^{-1}\left(\frac{101}{9}-20\ln{2}\right) \\
\label{eq:rad}
\delta(\Delta\nu_{\rm r-r}) &=& 
E_F\frac{\alpha(Z\alpha)^2}{\pi}\frac{m_e}{m_\mu}\ln(Z\alpha)^{-1}
\left(-\frac{431}{90} +  \frac{32}{3}\ln{2}  + Z^2  \right).
\eea
Numerically, these contributions give 
$-0.034 + 0.047 = 0.013\, {\rm kHz}$.  
Previous incomplete calculations~\cite{Kinoshita-talk}\cite{Nio-thesis} 
suggested a contribution of $-0.263(60)\, {\rm kHz}$.  The main result of this
paper is to show that in fact these contributions are not as large as the 
previous estimates.  
Remaining theoretical uncertainty is 
dominated by terms of 
order $E_F(Z\alpha)^3(m_e/m_\mu)\ln(m_\mu/m_e)$ 
($\sim 0.06\, {\rm kHz}$);
$E_F(Z\alpha)^3(m_e/m_\mu)$ ($\sim 0.03\, {\rm kHz}$); and  
$E_F\alpha(Z\alpha)^2(m_e/m_\mu)$ ($\sim 0.03\, {\rm kHz}$).  
A discussion of the error due to still uncalculated terms is
given at the end of the paper.

Including the complete results of Eqs.(\ref{eq:rec}),(\ref{eq:rad})
does not significantly alter the theoretical prediction for the 
HFS in physical units, which is in good agreement with the experimental value, 
Eq.(\ref{eq:expt}).  Here we simply quote Ref.~\cite{Mohr}(Eq.(D14)):
\be 
\Delta\nu({\rm Mu})_{\rm theory} = 4\,463\,302.67(27)\, {\rm kHz}
\,\,[6.1\times 10^{-8}],
\ee
where the error is due mainly to the measured value of 
$m_\mu/m_e$~\cite{error}.  
Likewise, the HFS 
determination of $\alpha$ is not significantly changed~\cite{Mohr}.
However, the new results in Eqs.(\ref{eq:rec}),(\ref{eq:rad}) represent 
a fractional shift of $6.2\times 10^{-8}$ in the HFS, and hence also
in the HFS determination of the mass ratio $m_\mu/m_e$.  
A recoil term of order $E_F(Z\alpha)^2(m_e/m_\mu)^2$
, which was not included in $\Delta\nu_{\rm rec}$ of Ref.\cite{Mohr},
contributes an additional 
$0.065(6)\,{\rm kHz}$ or $1.5\times 10^{-8}$\cite{Pachucki}.  
The mass ratio is then shifted from the value in Ref.~\cite{Mohr}(Eq.(161))
to become:
\be 
\label{eq:MU}
\left(m_\mu/m_e\right)[\Delta\nu({\rm Mu})] = 206.768\,2817(33)(24)(16)
\,\,[2.1\times 10^{-8}],
\ee
with the errors arising from uncertainty in $\Delta\nu_{\rm theory}$ due to uncalculated terms, from $\Delta\nu_{\rm expt.}$, and from the value 
of $\alpha$, respectively. 
This represents a shift of $2.5\sigma$ 
(in terms of the previous error $3.1\times 10^{-8}$),
and a $30\%$ reduction in error.

The positronium ($e^+e^-$)
HFS has also been measured precisely, though at present
its interest is for testing our knowledge of QED bound states, as
opposed to determining fundamental constants.   
The two most precise values are due to Mills and Bearman ($\Delta\nu(P)_{\rm expt.\,1}$, Ref.~\cite{Mills}) and Ritter {\it et. al.} ($\Delta\nu(P)_{\rm expt.\,2}$, Ref.~\cite{Ritter}):
\bea
\label{eq:expt1}
\Delta\nu({\rm Ps})_{\rm expt.\,1} &=& 203\,387.5(1.6)\, {\rm MHz} 
\,\,[7.9\times 10^{-6}]\\
\label{eq:expt2}
\Delta\nu({\rm Ps})_{\rm expt.\,2} &=& 203\,389.10(74)\, {\rm MHz} 
\,\,[3.6\times 10^{-6}].
\eea
The theoretical expression is:
\bea
\Delta\nu({\rm Ps})_{\rm theory} &=& m_e\alpha^4
	\left(C_0 + C_1\frac{\alpha}{\pi}+C_{21}\alpha^2\ln\alpha^{-1}
		+C_{20}\alpha^2 
\right. \nl &&\left.
 + C_{32}\frac{\alpha^3}{\pi}\ln^2\alpha^{-1}
		+C_{31}\frac{\alpha^3}{\pi}\ln\alpha^{-1}
		+C_{30}\frac{\alpha^3}{\pi} +\order(\alpha^4)
	\right).
\eea
Including the known terms through $C_{20}$~\cite{Czarnecki}
yields
$\Delta\nu({\rm Ps})_{\alpha^2} = 203\,392.93\, {\rm MHz}$. 
Coefficient $C_{32}=-7/8$ has been known for some time~\cite{Karshenboim},  
and in this
paper we calculate: 
\be
\label{eq:ann}
C_{31}=217/90-17\ln{2}/3.
\ee    
$C_{32}$ and $C_{31}$
contribute $-0.91\, {\rm MHZ}$ and $-0.32\, {\rm MHz}$ to 
$\Delta\nu({\rm Ps})_{\rm theory}$, respectively, bringing the theoretical 
prediction to:
\be
\Delta\nu({\rm Ps})_{\rm theory} = 203\,391.70(20)\, {\rm MHz}
\,\,[1.0\times 10^{-6}].
\ee
The uncertainty of $0.20\,{\rm MHz}$ corresponds to a coefficient $C_{30}\approx 4$.  For comparison, the numerical values of
the other coefficients are: $C_{0}=0.58$, $C_1=-1.24$, $C_{21}=0.21$,
$C_{20}=-0.39$, $C_{32}=-0.88$, $C_{31}=-1.52$. 
The discrepancy with experiment
is significant: $2.5\sigma $ and $3.5\sigma $ for 
Eqs.(\ref{eq:expt1}) and (\ref{eq:expt2}) respectively.  
As with the orthopositronium lifetime~\cite{Sapirstein-ops}\cite{ops},
a true disagreement between experiment and the predictions of QED would have
important consequences.  

The calculation is done in the framework of an effective 
quantum mechanical Hamiltonian theory~\cite{ops}, 
taking inputs from relativistic QED field theory and from (non-relativistic)
NRQED field theory~\cite{Caswell-Lepage}.  
The results to be derived for muonium can be translated directly to
positronium by taking $m_\mu\to m_e$, and including the additional 
contributions from virtual $e^+ e^-$ 
annihilation.

The Hamiltonian can be decomposed into the sum:
\be 
\label{eq:H}
H = H_0 + V_4 + V_5 + V_6 + V_7,
\ee
where $H_0$ is the unperturbed Hamiltonian for the Coulomb 
problem with reduced mass $m_r=m_em_\mu/(m_e+m_\mu)$:
\be
\label{eq:H0}
H_0 = \frac{p^2}{2m_r} -\frac{(Z\alpha)}{r}.
\ee
Potentials $V_4$, $V_5$, $V_6$ and $V_7$ give contributions 
to the energy of order $m\alpha^4$, $m\alpha^5$, etc. 
Since non-HFS operators will affect the HFS only in second- or 
higher-order perturbation theory, it follows that  
only the HFS parts of potentials $V_6$ and $V_7$ are necessary.
Furthermore, any potential not contributing to $S$-states
(in first or second order perturbation theory) may be 
neglected.

We will write the potentials in terms of 
a list of standard operators: ($q\equiv l-k$) 
\bea
\langle l|{\cal O}_1|k\rangle &=& \frac{1}{m_r^2} \nl
{\cal O}_2 &=& \frac{1}{\pi(Z\alpha)m_r^2}p^i\left(\frac{p^2}{2m_r}-\frac{Z\alpha}{r}-E\right)\ln\frac{m_r/2}{\frac{p^2}{2m_r}-\frac{Z\alpha}{r}-E}\,p^i \nl
\langle l|{\cal O}_3|k\rangle &=& \frac{1}{m_r^2}\ln\frac{|q|}{m_r} \nl
\langle l|{\cal O}_4|k\rangle &=& \frac{1}{m_r^2}\frac{|l\times k|^2}{q^2} \nl
\langle l|{\cal O}_5|k\rangle &=& 
{\pi(Z\alpha)}\frac{|q|}{m_r} \nl
\langle l|{\cal O}_6|k\rangle &=& \frac{q^2}{m_r^2}\ln\frac{|q|}{m_r} \nl
{\cal O}_7 &=& \frac{1}{\pi(Z\alpha)}\frac{p^4}{m_r^3} \nl
\langle l|{\cal O}_8|k\rangle &=& \frac{1}{m_r^2}\frac{|l\times k|^2}{q^4} \nl
\langle l|{\cal O}_{9}|k\rangle &=& \frac{1}{m_r^2}\left(\sigma_e\cdot\sigma_\mu-\frac{3q\cdot\sigma_e q\cdot\sigma_\mu}{q^2}\right) 
\eea
Note that ${\cal O}_1 = \delta^3(r)/m_r^2$.

Potential $V_4$ is derived from tree-level 
NRQED diagrams with Fermi, Darwin, Kinetic, and Dipole 
vertices~\cite{Nio-1},
and contains the leading relativistic corrections:
\bea
\label{eq:V4}
V_4 &=& 4\pi(Z\alpha)\left\{
\frac{c_F^e c_F}{6}^\mu\frac{m_r^2}{m_em_\mu}\left[\sig{\cal O}_1 +\frac{1}{2}{\cal O}_{9}\right]
 + \frac{1}{8}\left(c_D^e\frac{m_r^2}{m_e^2}+c_D^\mu\frac{m_r^2}{m_\mu^2}\right){\cal O}_1 \right. \nl
&&\left.
	-\frac{1}{32}\left(\frac{m_r^3}{m_e^3}+\frac{m_r^3}{m_\mu^3}\right){\cal O}_7 
	-\frac{m_r^2}{m_em_\mu}{\cal O}_8 \right\}.
\eea
Renormalization constants $c_F \approx c_D = 1 + \order(\alpha)$ 
are tabulated below.
$V_5$ gives the leading radiative corrections:
\bea
\label{eq:V5}
V_5 &=& \frac{2\alpha(Z\alpha)}{3}\left(\frac{m_r^2}{m_e^2}+2Z\frac{m_r^2}{m_em_\mu}+Z^2\frac{m_r^2}{m_\mu^2}\right){\cal O}_2 
	+\frac{14(Z\alpha)^2}{3}\frac{m_r^2}{m_em_\mu}{\cal O}_3 \nl
&&	+\left\{-\frac{4\alpha(Z\alpha)}{3}
		\left(\frac{m_r^2}{m_e^2}\ln\frac{m_r}{m_e}
		+Z^2\frac{m_r^2}{m_\mu^2}\ln\frac{m_r}{m_\mu}\right)\right. \nl
&& \left. + (Z\alpha)^2\frac{m_r^2}{m_em_\mu}\left[-\frac{2}{m_\mu^2-m_e^2}\left(m_\mu^2\ln\frac{m_e}{m_r}-m_e^2\ln\frac{m_\mu}{m_r}\right) +\frac{20}{9}\right.\right.\nl
&&\left.\left. -\frac{2m_em_\mu}{m_\mu^2-m_e^2}\ln\frac{m_\mu}{m_e}\sig\right]
 -\frac{4\alpha(Z\alpha)}{15}\frac{m_r^2}{m_e^2}\right\}{\cal O}_1.
\eea
Using first order perturbation theory, potentials $V_4$ and $V_5$ correctly reproduce the complete $S$-state energy spectrum 
through $\order(m\alpha^5)$\cite{ops}~\cite{Sapirstein-Yennie}.
The contribution from muon vacuum polarization is not relevant to
our analysis, and has been excluded from $V_5$. 

For $V_6$, only HFS terms are necessary.  These again are taken
directly from NRQED diagrams:  
\bea
\label{eq:V6}
V_6 &=& 4\pi(Z\alpha)
	\frac{\sig}{m_em_\mu}\left\{ \left[\frac{m_r^2}{m_em_\mu}
	\left(
	\frac{c_S^ec_S^{\mu}}{48}
	+\frac{ c_F^ec_F^\mu}{6}
	-\frac{c_F^ec_S^\mu+c_S^ec_F^\mu}{12}
	\right) \right.\right.\nl
&&\left.\left.
	-\frac{1}{24}\left(c_{p^\prime p}^ec_F^\mu\frac{m_r^2}{m_e^2}
		+c_F^ec_{p^\prime p}^\mu\frac{m_r^2}{m_\mu^2} \right)
	\right]{\cal O}_4 \right. \nl
&&\left.
	-\frac{1}{48}\left[c_S^ec_F^\mu\frac{m_r}{m_e}
		+c_F^ec_S^\mu\frac{m_r}{m_\mu} \right]{\cal O}_5
	\right\}.
\eea
Spin-Orbit, retardation, Time-Derivative, $p^\prime p$, and Seagull
interactions have been included.  
Additional local operator terms, 
of the form $-\nabla^2\delta^3(r)$  and $\{p^2,\delta^3(r)\}$
are not shown explicitly; these analytic terms do not generate
factors of $\ln\alpha$, 
and so are not relevant to the present analysis~\cite{Labelle}.  

The necessary renormalization
constants have already been calculated\cite{Nio-1}\cite{constants}: 
\be
c_F^e = 1 + a_e, \quad
c_D^e = 1 + \frac{8\alpha}{3\pi}\left(-\frac{3}{8}+\frac{5}{6}\right) + 2a_e, 
\quad
c_S^e = 1 + 2 a_e, \quad
c_{p^\prime p}^e = a_e.
\ee
Here $a_e=\alpha/2\pi + \order(\alpha^2)$ is the electron anomalous magnetic moment. For $c^\mu$, $m_\mu$ and $Z^2\alpha$ are substituted 
for $m_e$ and $\alpha$.

Potential $V_7$ has no non-instantaneous 
HFS contribution coming from photon momenta 
$q\approx m\alpha^2$, a consequence of the fact that
spin-dependent M1 multipole 
transitions vanish in the absence of relativistic effects,
and are therefore suppressed. 
The remaining instantaneous part of $V_7$, from momenta $q\approx m\alpha$, 
is fully determined by requiring that the Hamiltonian correctly 
reproduce the 
low-momentum expansion of the 1-loop photon-exchange scattering amplitude.
Introducing a photon mass $\lambda$, 
and ultraviolet cutoff $\Lambda$ on photon momenta,
the effective Hamiltonian (without $V_7$) gives~\cite{effective}:
\be
\left[\frac{2\pi(Z\alpha)}{3m_em_\mu}\sig\right]\frac{(Z\alpha)}{\pi}\frac{q^2}{m_em_\mu}\left(-\frac{2}{3}\ln\frac{\Lambda}{\lambda} + \cdots\right),
\ee
where again analytic terms are not shown.
The corresponding 
QED amplitude is (Fig. 1):
\be
\left[\frac{2\pi(Z\alpha)}{3m_em_\mu}\sig\right]\frac{(Z\alpha)}{\pi}\frac{q^2}{m_em_\mu}\left(-\frac{2}{3}\ln\frac{\Lambda}{\lambda}
+\frac{1}{4}\log\frac{q}{\Lambda} + \dots\right).
\ee
This result has been checked both in QED Feynman gauge, 
and in NRQED Coulomb gauge\cite{vertex}.  
Requiring the effective theory to match QED implies that 
\be
\label{eq:V7}
V_7 = \frac{(Z\alpha)^2}{6}\frac{m_r^2}{m_e^2m_\mu^2}\sig{\cal O}_6.
\ee
Contributions to $V_7$ having a
dependence on $m_e$, $m_\mu$, $\alpha$ and $Z$ different from Eq.(\ref{eq:V7})
 are ruled out by noticing that: (i) The non-recoil contributions are
already present in $V_4$, $V_5$ and $V_6$ (as we will soon verify), so that
$V_7$ contains no non-recoil piece;
(ii) Masses can enter only as inverse powers $1/m_e$ and  $1/m_\mu$,
and in particular not as $1/(m_e+m_\mu)$. This latter result
can be seen clearly using time ordered perturbation theory in NRQED:
the NRQED vertices are all homogeneous in the masses, leaving only
the energy denominators to consider; however, the energy
denominators will all have the form $1/(|q| + p_1^2/m_e + p_2^2/m_\mu)$,
with photon momentum $q$ and particle momenta $p_1$, $p_2$.
(Contributions which are not simply iterations of lower-order potentials
must have at least one photon in each intermediate state.)
Such an expression, for $q\approx p_1\approx p_2\approx m\alpha$, can be expanded in powers 
of $p_1^2/m_e|q|$, $p_2^2/m_\mu|q|$---again homogeneous in the
masses.  Using (i) and (ii), the only possible parameter dependence 
which is symmetric in $m_e$ and $m_\mu$ is that of Eq.(\ref{eq:V7}). 

Having completed the specification of the Hamiltonian, Eqs.(\ref{eq:H}),(\ref{eq:H0}),(\ref{eq:V4}),(\ref{eq:V5}),(\ref{eq:V6}),(\ref{eq:V7}), we now 
use the usual expressions from Rayleigh-Schr\"{o}dinger perturbation theory
to solve for the energy shift:
\be
\Delta E = \left\langle V_6 + V_7 \right\rangle + \left\langle (V_4+V_5) \tilde{G} (V_4+V_5)\right\rangle 
	+\left\langle V_4 \right\rangle \left\langle \frac{\partial V_5}{\partial E} \right\rangle ,
\ee
where $\tilde{G}$ is the Coulomb Green's function with ground state pole
removed, and $\langle V \rangle$
is the expectation value of $V$ in the ground state of the unperturbed 
$H_0$, Eq.(\ref{eq:H0}). 
The logarithmic contributions of the necessary matrix elements 
are:
\be
\label{eq:first-order}
\frac{\langle {\cal O}_i \rangle}{\langle \delta^3(r)\rangle} \to
(Z\alpha)^2\ln(Z\alpha)^{-1}
\left\{
\begin{array}{c}
2,\quad i=4 \\
8,\quad i=5 \\
12, \quad i=6 \\
\end{array}
\right.
\ee
\be
\label{eq:second-order}
\left(2\frac{\langle {\cal O}_i\tilde{G}\delta^3(r)\rangle}{\langle \delta^3(r)\rangle} +  \langle\frac{\partial {\cal O}_i}{\partial E}\rangle\right) 
\to \frac{(Z\alpha)}{\pi}\ln(Z\alpha)^{-1} \times 
\left\{
\begin{array}{c}
-2,\quad i=1 \\
-4\ln(Z\alpha)^{-1} +6 -8\ln{2},\quad i=2 \\
\ln(Z\alpha)^{-1} + 1 - 2\ln{2},\quad i=3 \\
-16,\quad i=7 \\
-1,\quad i=8 \\
\end{array}
\right.
\ee
\be
\frac{\langle {\cal O}_{9} \tilde{G} {\cal O}_{9} \rangle}
	{\langle \delta^3(r)\rangle} 
	\to \frac{10}{m_r^2} \frac{(Z\alpha)}{\pi}\ln(Z\alpha)^{-1} ,
\ee
where the arrows signify that only logarithmic corrections, and
in the case of $\langle{\cal O}_{9}\tilde{G}{\cal O}_{9}\rangle$, only the HFS part, are shown~\cite{UV}. 
The pure recoil result for the HFS at order 
$E_F(Z\alpha)^3(m_e/m_\mu)$ 
contains the previously known $\ln^2(Z\alpha)$ and 
$\ln(Z\alpha)\ln(m_\mu/m_e)$ contributions~\cite{Karshenboim}\cite{Nio-1}
\cite{recoil-log}; 
the new $\ln(Z\alpha)$ term is shown in 
Eq.(\ref{eq:rec})~\cite{difference}.   
For radiative corrections at order $E_F\alpha(Z\alpha)^2$,
the non-recoil $\ln^2(Z\alpha)$ and $\ln(Z\alpha)$ terms, and the 
recoil $(m_e/m_\mu)\ln^2(Z\alpha)$ term~\cite{Karshenboim}, 
agree with previous calculations. 
A part of the radiative-recoil single-logarithm corresponding to reduced mass 
factor $m_r^2/m_em_\mu \approx (1-2m_e/m_\mu)$ was included 
previously~\cite{Nio-thesis}\cite{Kinoshita-talk};
the complete contribution is given in Eq.(\ref{eq:rad}).  
Numerical values are summarized in Table~I.

For positronium, there are additional interactions due to 
virtual annihilation of the electron and positron.  
The hard annihilation process is described by local operators,
which by themselves cannot generate nonanalytic factors.
So, for $\ln\alpha$ 
contributions, only second order perturbations involving 
$V_4$ and $V_5$ need be considered: 
\be
\delta V_4 = \frac{\pi\alpha}{2}\left(\frac{3}{4}+\frac{\sigma_e\cdot\sigma_\mu}{4}\right){\cal O}_1
\ee
\be
\delta V_5 = \alpha^2\left[\left(-\frac{22}{9}\right)\left(\frac{3}{4}+\frac{\sigma_e\cdot\sigma_\mu}{4}\right) 
	 + \left(-1+\ln{2}\right)
		\left(\frac{1}{4}-\frac{\sigma_e\cdot\sigma_\mu}{4}\right)
		\right]{\cal O}_1 .
\ee
$\delta V_4$ gives the leading contribution from 1-photon annihilation. 
The first and second terms of $\delta V_5$ come from 
radiative corrections to $\delta V_4$, and 
from 2-photon virtual annihilation, respectively. 
$\order(m\alpha^7\ln\alpha)$ contributions from these annihilation
operators are:
\bea
\Delta\nu_{\rm ann.}(m_e\alpha^7\ln\alpha) &=& m_e\frac{\alpha^7}{\pi}\ln\alpha^{-1}
\left[ -\frac{3}{8}\ln\alpha^{-1} +\frac{2261}{1080}-3\ln{2}
\right].
\eea
The non-annihilation contributions for positronium are obtained by taking the
limit $m_\mu\to m_e$ in the muonium analysis (making no expansion
in $m_e/m_\mu$); the combined result is given in Eq.(\ref{eq:ann}).

The previously most significant sources of error in the muonium HFS 
were 
$\Delta\nu_{\rm r-r}$ ($0.104\,{\rm kHz}$) and 
$\Delta\nu_{\rm rec}$ ($0.060\,{\rm kHz}$)\cite{Mohr}; all other 
uncertainties are estimated below 
$0.010\,{\rm kHz}$~\cite{nu-rad}.  
By calculating the $\order(E_F\alpha^3(m_e/m_\mu)\ln\alpha)$ contribution
to $\Delta\nu_{\rm r-r}$, the uncertainty in this quantity 
should be reduced by a factor $\sim\ln\alpha^{-1}\approx 5$;
in fact, since there are still uncalculated terms at
$\order(E_F\alpha^2(Z\alpha)(m_e/m_\mu)\ln(m_\mu/m_e))$~\cite{Eides-log} and
$\order(E_F\alpha(Z\alpha)^2m_e/m_\mu)$~\cite{constant},
we take this uncertainty as $0.040\,{\rm kHz}$. 
The uncertainty in $\Delta\nu_{\rm rec}$ should remain 
approximately the same, since it is dominated by the
still uncalculated terms of order 
$\order(E_F(Z\alpha)^3(m_e/m_\mu)\ln(m_\mu/m_e))$~\cite{log}
and $\order(E_F(Z\alpha)^3(m_e/m_\mu))$~\cite{constant}.  
Thus we take $0.070\,{\rm kHz}$
as an estimate of the total remaining theoretical error. 

In the final stages of the calculation, I 
received word from K. Melnikov and A. Yelkhovsky 
that they have also performed the calculation of $\alpha^3\ln\alpha$ terms,  
in a dimensional regularization approach~\cite{Melnikov}.
After a detailed comparison,
we agree fully on the contributions in both muonium and positronium.
The agreement found in different formalisms in two independent
calculations lends strong support to the correctness of the
results. 

This work was motivated in part by, and is an extension of, 
Ref.~\cite{ops}.  Many ideas used in the calculation
originated with G.~P. Lepage, who I thank for continued insights
and encouragement during the present work.  Thanks are also due
to  P. Labelle, and to K. Melnikov and A. Yelkhovsky 
for useful conversations.  This work was supported by a grant from the
National Science Foundation.


\begin{table}
\begin{center}
\begin{tabular}{cll}
$\times{E_F}\frac{m_e}{m_\mu}$ & Ref.~\cite{Mohr}  (kHz) & present paper (kHz) \\
\hline
$(Z\alpha)^3\ln^2{(Z\alpha)}$ & $-0.043$ \\
$(Z\alpha)^3\ln{(Z\alpha)}\ln({m_\mu}/{m_e})$ & $-0.210$ \\
$(Z\alpha)^3\ln{(Z\alpha)}$ & $-0.257(*)$ & $-0.034$ \\
$(Z\alpha)^3\ln({m_\mu}/{m_e})$ & --- & $-0.035$ (*)\cite{log} \\
$(Z\alpha)^3$ & $0.107(30)$ \\
\hline
$\alpha(Z\alpha)^2\ln^2{(Z\alpha)}$ & $0.344$ \\
$\alpha(Z\alpha)^2\ln{(Z\alpha)}$ & $-0.008$ (*) & $0.034$ \\
$\alpha(Z\alpha)^2$ & $-0.107(30)$ \\
$Z^2\alpha(Z\alpha)^2\ln{(Z\alpha)}$ & --- & $0.013$ \\
\hline
$\alpha^2(Z\alpha)\ln^3({m_\mu}/{m_e})$ & $-0.055$ \\
$\alpha^2(Z\alpha)\ln^2({m_\mu}/{m_e})$ & $0.010$ \\
$\alpha^2(Z\alpha)\ln({m_\mu}/{m_e})$ & $0.009$ (*) \\
$\alpha^2(Z\alpha)$ & --- \\
\end{tabular}
\caption{ Contributions of order $E_F\alpha^3(m_e/m_\mu)$ to the muonium HFS.
The second column lists the contributions used in Ref.[3]; 
the third column gives new or modified values from the present paper. 
Asterisks denote partial results.
}
\end{center}
\label{table: status}
\end{table}

%


\begin{thebibliography}{99}
\bibitem{error}
The fractional error of 
$\Delta\nu_{\rm theory}$ is $\sim 1/2$ that of 
$m_\mu/m_e$, due to the manner in which the primary 
data of Ref.~\cite{Liu} was used.
\bibitem{constants}
The cutoff-dependent part of $c_D$ from Ref.\cite{Nio-1}, Eq.(56), has been 
absorbed into $V_5$.
\bibitem{effective}
The analysis is similar to the calculation of $c_{q^2}$ in 
section C of Ref.~\cite{Nio-1}.  The amplitude is defined using 
non-relativistic normalization for the external particles---see
Note~[7] of Ref.~\cite{ops}.
\bibitem{vertex}
For the NRQED case, 
a spin-dependent 
``seagull'' interaction with two magnetic photons is necessary.  
The Feynman rule is:
$
{ie^2}/{8m^2} (p_0 - q_0) \epsilon_{ijk} \sigma_i 
$,
where $(p,j)$ and $(q,k)$ are the two incoming photon momenta and 
polarization directions.  This vertex, 
coming from the $\sigma\cdot E\times D$ part of the NRQED Lagrangian,
was not necessary in the non-recoil calculation performed in 
Ref.\cite{Nio-1}.
\bibitem{UV}
Logarithmic contributions are independent of UV cutoff---see
the discussion in Ref.\cite{ops}.
\end{thebibliography}
\end{document}